\documentstyle[12pt,preprint,aps]{revtex}
\begin{document}
\draft
\title{
Quantum phase transitions in the $J-J'$ Heisenberg \\
and XY spin-$\frac{1}{2}$ antiferromagnets on square lattice: \\
Finite-size scaling analysis }
\author{Piotr Tomczak}
\address{
Physics Department, Adam Mickiewicz University,\\
Umultowska 85, 61-614 Pozna\'{n}, Poland\\
}
\author{Johannes Richter}
\address{
Institut f\"ur Theoretische Physik, Otto-von-Guericke Universit\"at,
Magdeburg,\\
P.O.B. 4120, 39016 Magdeburg, Germany\\
}

\maketitle

\begin{abstract}
We investigate the critical parameters of an order-disorder 
quantum phase transitions in the spin-$\frac{1}{2}$  $J-J'$ Heisenberg
and XY antiferromagnets on square lattice.  
Basing on the excitation gaps calculated by exact diagonalization technique for systems
up to 32 spins and finite-size scaling analysis we estimate the critical 
couplings and exponents of the correlation length for both models.
Our analysis confirms the universal critical behavior of these quantum
phase transitions:
They belong to 3D O(3) and 3D O(2) universality classes, respectively.
\end{abstract}





The equivalence of the critical behavior of  $D$ dimensional quantum 
spin systems and $D+1$ dimensional
classical spin systems is well recognized. This idea,
combined with finite-size scaling was used previously many times
 to discuss the critical properties of
infinite spin systems, see e.g., \cite{{HaBa},{Ham}} and \cite{Bar} for a review. 
However, these investigations were strongly limited with
respect to the system size.
Due to the recent advances in computer technology it is possible to treat
bigger systems, e.g., up to 36 sites for spin $\frac{1}{2}$, and consequently
to extract their critical properties using finite-size scaling 
method \cite{Ha1,Ha2}. Our aim is to present in this paper the results 
of such an investigation 
of critical parameters (coupling and exponents of correlation length)
for the $J-J'$ Heisenberg and XY spin-$\frac{1}{2}$ antiferromagnets on square lattice.

The Hamiltonian of the model whose critical behavior is examined is given by
\begin{equation}
\label{eq:1}  
H=J\sum_{<i,j>}\roarrow{S}_{i}\cdot \roarrow{S}_{j}
 +J'\sum_{<k,l>}\roarrow{S}_{k}\cdot \roarrow{S}_{l}. 
\end{equation}
The first sum, denoted by $<i,j>$, runs over pairs of nearest-neighbors 
on the square lattice connected by thin bonds (see Fig.1), whereas 
the second one, denoted by $<k,l>$ --- over nearest-neighbors connected by thick bonds.
In the case of Heisenberg spin system three Pauli matrices are 
included into scalar product in Eq. (\ref{eq:1}), in the case of XY system --- only two.
The model represents an antiferromagnet, i.e., both couplings are positive and 
additionally, $J' \ge J$. 
Clearly, what one can see here is the competition between long-range N\'eel order 
and the tendency to the formation of local singlets of two neighboring spins, coupled via $J'$.
In limiting cases this model
reduces to the long-range ordered Heisenberg antiferromagnet on square lattice for $J=J'$
on one hand and on the other ---
to  disjoint singlets (no staggered magnetization) for $J'/J \to \infty$.
At some $(J'/J)$ there exists a quantum phase transition between the gapless Ne\'el phase
and a gapped 'singlet' phase (quantum paramagnet). 
The properties of $J-J'$ Heisenberg model on the square lattice were first examined
by series expansion (SE) \cite{Si}  and more recently by renormalized spin wave (RSW)
approach  \cite{Iv}, exact diagonalization (ED) and coupled cluster
method (CCM) \cite{Kr} in order to observe the interplay between
the local singlet formation tendency and the long-range N\'eel order.  
Although, in general, all those methods predict the existence of quantum phase
transition, they differ drastically in the estimation of the critical coupling.
Furthermore, there was only one attempt \cite{Si} to find critical exponents,
however the error of this estimation was rather large. 

Let  us now rewrite the Hamiltonian in such a form that the two above tendencies will be 
seen more clearly:
\begin{equation}
\label{eq:2}  
H=g\big(\sum_{<i,j>} \roarrow{S}_{i}\cdot \roarrow{S}_{j}
 +\frac{1}{g^2}\sum _{<k,l>}\roarrow{S}_{k}\cdot \roarrow{S}_{l}\big), 
\end{equation}
$0<g\le1$. Note that in Eq. (\ref{eq:2}) 
one has the same summation specification as in
Eq. (\ref{eq:1}). The coupling constant $g$ determines the relevant
energy scale in the model under consideration and  
the term $1/g^2=\lambda$ is analogous to an inverse temperature.

In what follows the estimation of the critical value of $g_c$ and critical exponent
$1/\nu$ for the Heisenberg and XY Hamiltonians will be described.
At the beginning one finds by the exact diagonalization (Lanczos algorithm) 
the spin-gap, defined as $\Delta = E_1 - E_0$ ($E_{0(1)}$ is the
lowest energy in the $S^z=0(1)$ sector)  dependence
vs. $\lambda = 1/g^2$ for the Hamiltonian (\ref{eq:2}) on a square lattice for a sequence
of finite systems. Note that all the systems, being elements of the
sequence should be invariant under the same symmetry operations;
in the opposite case it is not possible to find a proper scaling.
In the system under examination one has only three such  systems: with $N = 8, 18, 32$ spins. 
Two of them are shown in Fig. 1, the third  one has the same shape.

The spin-gap, multiplied by the linear dimension of the system, $\Delta\sqrt N$, allows one to
find the pseudo-critical points\cite{Bar}: for the Heisenberg and XY systems
one has three such points, collected in Table 1 (Heisenberg) and 2 (XY). 
Next one should extrapolate the sequence of pseudo-critical points to infinity.
However, it is not possible to employ here any algorithm improving the
convergence of the finite-lattice data sequence the series is extremely short.
Therefore we find it more accurate to use graphical methods
to find the critical value of $g$. The
sequences of pseudo-critical points for Heisenberg and XY system 
plotted vs. $1/L^4$  are shown in Fig. 2. The estimate
of 1/$g_c^2=2.46(2)$ for the infinite Heisenberg system should be compared to the value
2.56 obtained by SE \cite{Si}, to the value of 3.16 obtained by 
CCM method, to the value of 2.45 from ED \cite{Kr}
and finally to the value of 5.0 \cite{Iv} from RSW approach. 
The extrapolated value of $1/g_c^2 = 4.56(2)$  for $anisotropic$ XY system
is higher, as one should expect, than that for Heisenberg system
since anisotropy act aginst the singlet formation \cite{De}. 

The critical exponent for the correlation length, $\nu$, may be estimated 
from the behavior of the Callan-Symanzik $\beta$-function \cite{Ha79,Bar,HaBa}
\begin{equation}
\label{eq:3}
\beta(\lambda)/g = \frac{d}{dg} \ln [g \Delta(\lambda)],
\end{equation}    
which, calculated for a finite system of linear size $L$ in a pseudo-critical point,
 scales as
\begin{equation}
\label{eq:4}
\beta(\lambda_c,L) \sim L^{-1/\nu}.
\end{equation} 
Usually, in order to see this scaling behavior, one takes into
account a sequence of $\beta(L_{i})/\beta(L_{i-1})$ values calculated 
at the pseudo-critical points for some
values of $L_i$, such that $L_{i}=L_{i-1}+1$. Expanding
\begin{equation}
\label{eq:5}
\beta(L_{i})/\beta(L_{i-1}) \sim (1+1/L_{i})^{-1/\nu} 
\sim  1 - \frac{1}{\nu} \frac{1}{L_{i}} + ...
\end{equation}  
one finds a linear behavior of $(1-\beta(L_{i})/\beta(L_{i-1})$ vs. $1/L$
for all $i$. This is the 'linear' 
approximation. Note that the error is $O(1/L^2)$. However, if one
can not find a sequence of finite systems fulfilling $L_{i}=L_{i-1}+1$
(in our case $L_{i}=L_{i-1}+\sqrt2$) this approximation is rather crude because of 
the order of the error. This may be improved in the following way.
First, let us note that in the following expansion, for small $x$
\begin{equation}
\label{eq:6}
\ln\bigg(\frac{1+x}{1-x}\bigg)=2x + \frac{2}{3}x^3 + ...
\end{equation} 
the term $x^2$ is absent and the error is $O(x^3)$.
Second, let us put 
\begin{equation}
\label{eq:7}
x = \frac{L_{i}-L_{i-1}}{L_{i}+L_{i-1}},
\end{equation}   
and expand
\begin{equation}
\label{eq:8}
\ln[\beta(L_{i-1})/\beta(L_{i})] \sim  \frac{1}{\nu} \ln 
\bigg( \frac{1+x}{1-x} \bigg)
\sim   \frac{2}{\nu} \frac{(L_{i}-L_{i-1})}{(L_{i}+L_{i-1})} + ...
\end{equation}   
Consequently on has a linear dependence of $\ln[\beta(L_{i})/\beta(L_{i-1})]$ vs $1/L$:
\begin{equation}
\label{eq:9}
\ln[\beta(L_{i-1})/\beta(L_{i})] 
\sim   \frac{\sqrt2}{\nu} \frac{1}{\bar L_i} + ...,    
\end{equation}   
where $\bar L_i = (L_{i}+L_{i-1})/2$.
The main advantage of this approach is a small finite size error, what
in consequence enables one to examine smaller systems. However, there remains
yet another problem. It is possible to consider
scaling from two pseudo-critical points: (8-18) where $x=1/5$ and (18-32)
 --- where $x=1/7$.
In the third point the expansion \ref{eq:8} gives rather large
finite size correction $(x=1/3)$ and this point has to be excluded. 
Thus, one can ask whether it is possible to find a scaling
relation 
and to estimate critical exponent from two points only?
The answer is yes, but the final error will be larger.
Since the expansion (\ref{eq:8}) produces an error  $O(x^3)$, it
seems to be especially well suited to this purpose.
To test this approach we have extrapolated the $1/\nu$ exponent 
from the data for transverse Ising model on square lattice \cite{Ha2}
taking into account only two pseudo-critical points: (16-25)
and (25-36). The estimate of $1/\nu=1.586(7)$ obtained by simple
linear extrapolation from these two points should
be compared to the original one $1/\nu=1.591(1)$ \cite{Ha2},
extrapolated from four points.

The data collected in Table 1 and 2 enable one to calculate 
$\beta$-function from Eq. (\ref{eq:3}) and consequently 
$1/\nu$ from  Eq. (\ref{eq:9}) for each pseudo-critical point of $J-J'$ model;
their values for Heisenberg and XY Hamiltonians are listed in 
Table 3 and the graphical extrapolation is shown in Fig 3.
 The values of
critical exponents for the same universality classes
obtained by other authors are also displayed in Table 3.  

One should note that the error in the present approach is larger 
than in other finite-size scaling analysis, but as was mentioned
this is a result of the small number of systems with required symmetry,
see Fig. 1, and the error is still comparable with an error
resulting from extensive Monte-Carlo simulation \cite{{TIU},{Ch}}.

To conclude, we have presented the results of the 
investigation of  the critical parameters
for the quantum phase transitions in the spin-$\frac{1}{2}$ $J-J'$ Heisenberg
and XY antiferromagnets on square lattice.
The obtained values of the correlation length critical exponents strongly 
suggest that these transitions belong to the 
3D O(3) and 3D O(2) universality classes, respectively.  

{\bf Acknowledgments} The authors thank J\"org Schulenburg for 
the numerical assistance and Prof. Ryszard Ferchmin for reading
manuscript.  Support from the Polish Committee for
Scientific Research (Project No. 2 PO3B 046 14) and from the Deutsche
Forschungsgemeinschaft (Projects No. 436/POL17/5/01  and Ri 615/10-1)
are also acknowledged. 
Some of the calculations were performed at the Pozna\'n Supercomputer 
and Networking Center.

\begin{figure}
\label{pierw}
\caption{ 
The $J-J'$ model on square lattice. Finite
systems of 8 and 18 spins are marked by dashed lines.}
\end{figure}

\begin{figure}
\label{drug}
\caption{ 
Extrapolation of the finite-system pseudo-critical points as
a function of $1/L^4$. Open squares - Heisenberg model, filled squares - XY model.}
\end{figure}

\begin{figure}
\label{trze}
\caption{ 
Extrapolation of the finite-system estimates 
of the $1/\nu$ exponent a function of  $1/L$. 
Open squares - Heisenberg model, filled squares - XY model.}
\end{figure}

\begin{table}
\label{table1}
\caption{
Pseudo-critical points $\lambda_c$ calculated for two
Heisenberg systems of sizes given in the first column.
The values of the gap $\Delta$ and its derivative $\Delta'$
at the pseudo-critical point $\lambda_c$ are also listed. To find them
11 ED data points equally spaced around $ \lambda_c$ were fitted 
to the polynomial of 4-th order in $\lambda$ in the region 
of $\lambda_c \pm 0.005$. 5 digits are exact.}
\begin{tabular}{cccc}
System size &$\lambda_c$  & $\Delta$  & $\Delta'$ \\ \hline
 8          & 3.1166      &  4.14495  & 1.52350  \\         
18          &             &  2.76330  & 1.71055  \\ \hline
 8          & 2.9251      &  3.85980  & 1.45296  \\         
32          &             &  1.92990  & 1.81304  \\ \hline
18          & 2.7648      &  2.20467  & 1.44663  \\           
32          &             &  1.65350  & 1.62565  \\ \hline          
extrapolated& 2.46(2)     &           &          \\         
\end{tabular}
\end{table}

\begin{table}
\label{table2}
\caption{
Same as TABLE I, but for XY spin systems.}
\begin{tabular}{cccc}
System size &$\lambda_c$  &  $\Delta$  & $\Delta'$ \\ \hline
 8          & 4.9326      & 2.51653    & 0.786928 \\       
18          &             & 1.67768    & 0.848422 \\ \hline
 8          & 4.7705      & 2.39028    & 0.770058 \\       
32          &             & 1.19514    & 0.942901 \\ \hline
18          & 4.6444      & 1.44152    & 0.788049 \\       
32          &             & 1.08114    & 0.866564 \\ \hline  
extrapolated& 4.56(2)     &            &          \\       
\end{tabular}
\end{table}

\begin{table}
\label{table3}
\caption{
Values of critical exponent $1/\nu$ calculated for two pseudo-critical points
for finite Heisenberg and XY spin systems and subsequently
extrapolated to infinity. For comparison are also included values of the same exponents
calculated by other authors.}
\begin{tabular}{cccc}
System size       &Heisenberg O(3) & XY, O(2)  \\ \hline
 8-18             & 1.9841                  & 1.6220        \\
18-32             & 1.8323                 & 1.6014                \\ \hline   
extrapolated      & 1.44(4)                 & 1.55(4)               \\ \hline  
                  & 1.46(8)  \cite{TIU}     & 1.495(5)\cite{Le1}    \\
                  & 1.418(6) \cite{Ch}      & 1.49(1) \cite{Le2}    \\
                  &                         & 1.51(1) \cite{Go}     \\
\end{tabular}
\end{table}


\begin{references}

\bibitem{HaBa} C. J. Hamer and Michael N. Barber, J. Phys. A: Math. Gen. {\bf 13}, L169 (1980).
\bibitem{Ham} C. J. Hamer, J. Phys. A: Math. Gen. {\bf 15}, L675 (1982).
\bibitem{Bar} M. N. Barber, {\em Phase Transitions and Critical Phenomena} vol. 8, \\
eds. C. Domb  and J. L. Lebowitz, Academic, New York (1981).
\bibitem{Ha1} C. J. Hamer, Tobias H\"ovelborn and Michael Bachhuber, \\
 J. Phys. A: Math. Gen. {\bf 32}, 51 (1999). 
\bibitem{Ha2} C. J. Hamer, J. Phys. A: Math. Gen. {\bf 33}, 6683 (2000).
\bibitem{Si}  Rajiv R. P. Singh, Martin P. Gelfand and David A. Huse, \\  Phys. Rev. Lett {\bf 61}, 2484 (1988). 
\bibitem{Iv}  N. B. Ivanov, S. E. Kr\"uger and J. Richter, Phys. Rev. B {\bf 53}, 2633 (1996). 

\bibitem{De}  O. Derzhko, J. Richter, O. Zaburannyi, J. Phys. Condens. Matter {\bf 12}, 8661 (2000). 
\bibitem{Kr}  Sven E. Kr\"uger, Johannes Richter, J\"org Schulenburg,
Damian J. J. Farnell, \\ Raymond F. Bishop, Phys. Rev. B {\bf 61}, 14607 (2000).
\bibitem{Ha79}  C. J. Hamer, J. Kogut,  Phys. Rev. B {\bf 20}, 3859 (1979). 
\bibitem{TIU} M. Troyer, M. Imada, and K. Ueda, J. Phys. Soc. Jpn. {\bf 66}, 2957 (1997).
\bibitem{Ch}  K. Chen, A. M. Ferrenberg and D. P. Landau,   Phys. Rev. B {\bf 48}, 3249 (1993). 
\bibitem{Le1}  J. C. Le Guillou and J. Zinn-Justin,  Phys. Rev. B {\bf 21}, 3976 (1980). 
\bibitem{Le2}  J. C. Le Guillou and J. Zinn-Justin,  J. Physique Lett. {\bf 46}, L-137 (1985). 
\bibitem{Go}  A. P. Gottlob and M. Hasenbusch, J. Stat. Phys. {\bf 77}, 919 (1994). 
\end{references}
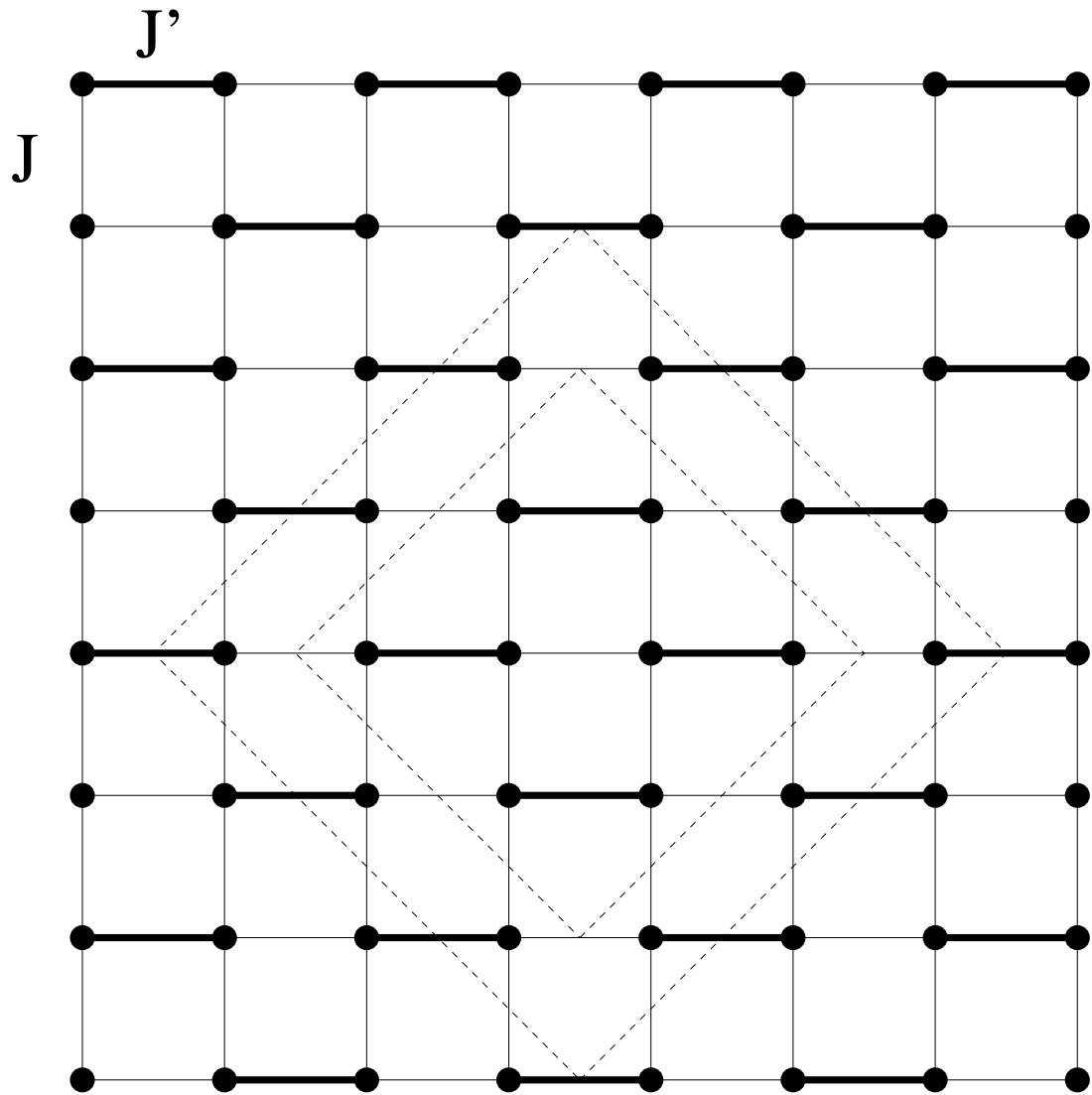
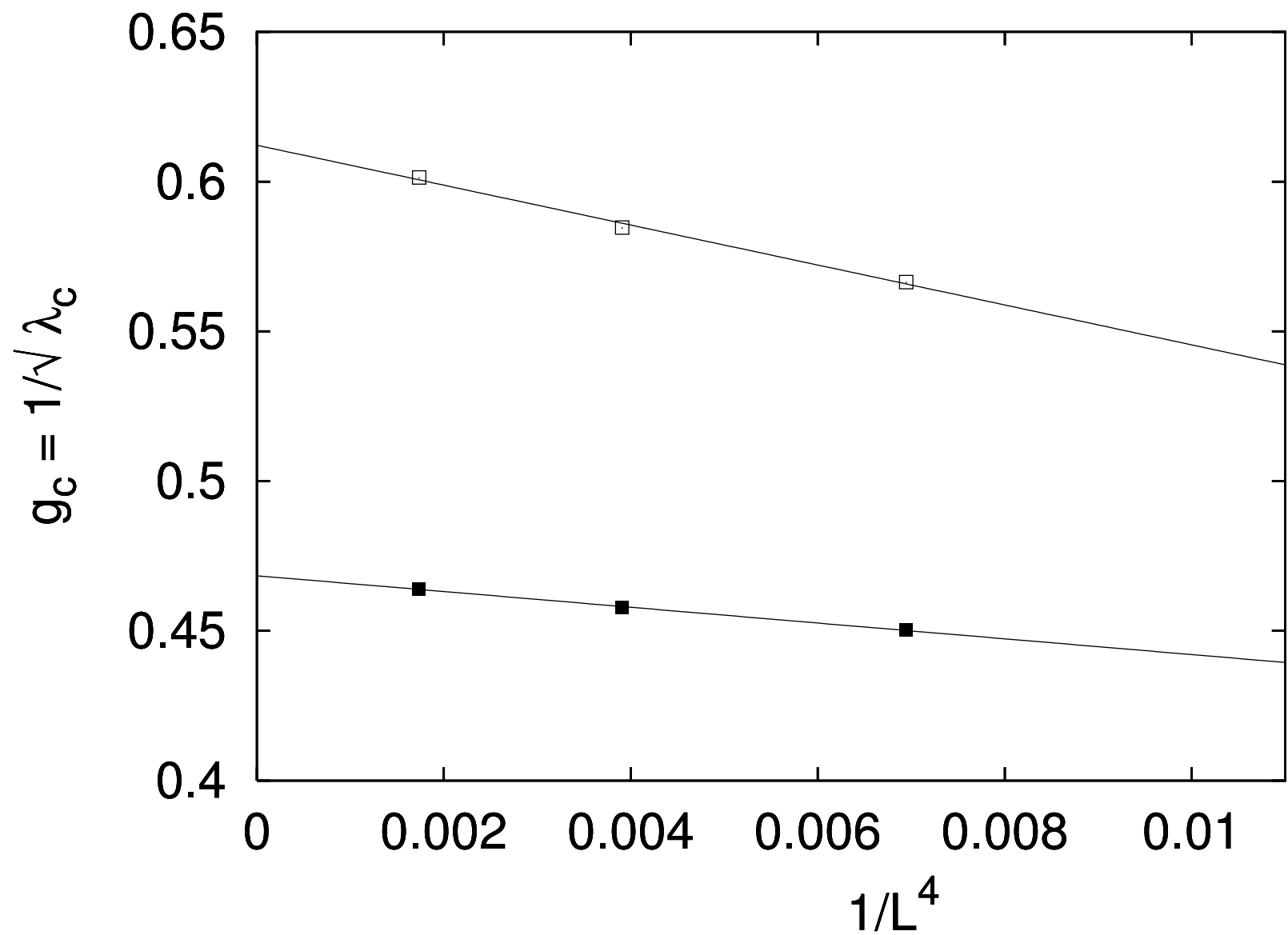
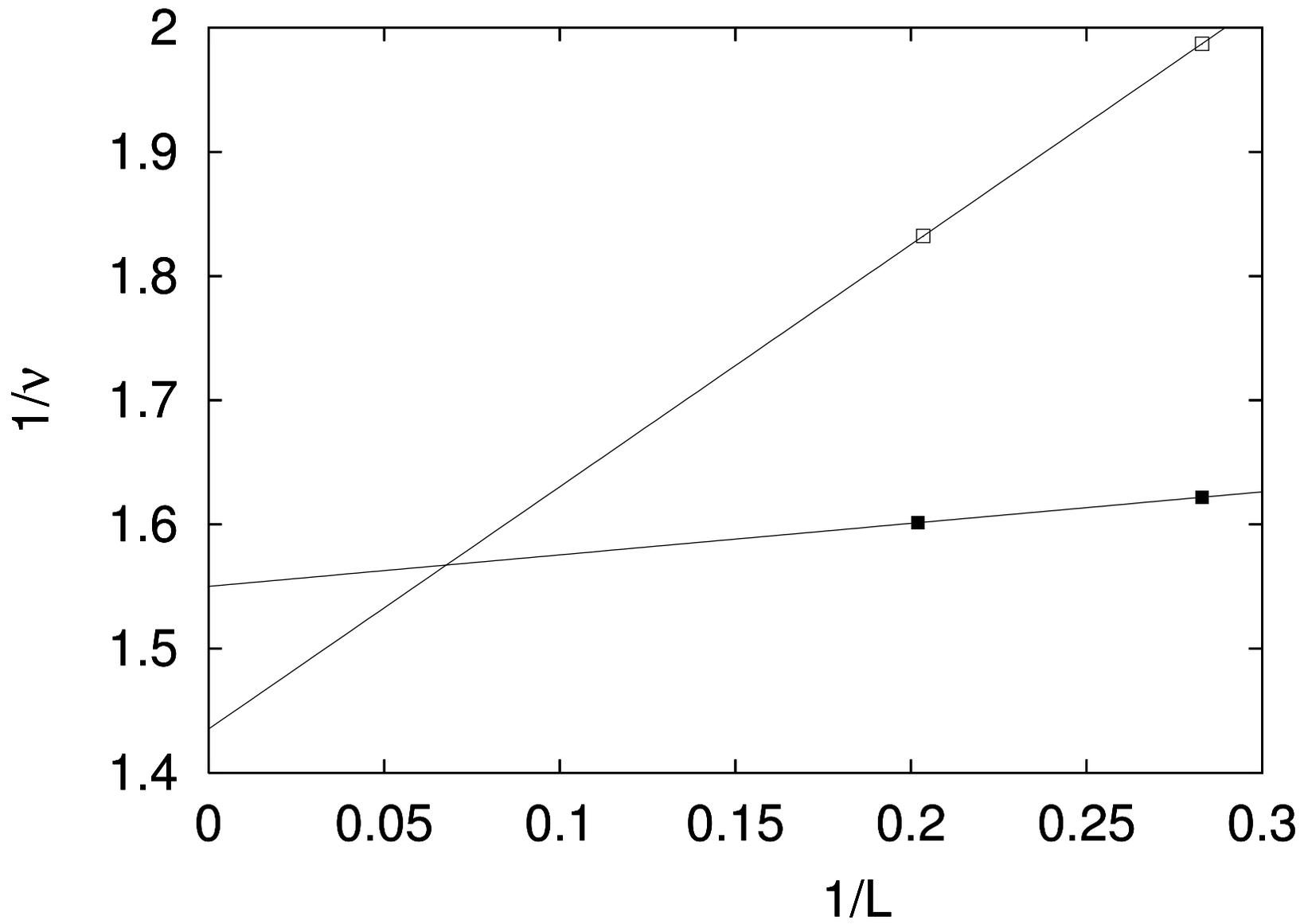
\end{document}